\documentclass[conference,hidelinks]{IEEEtran}
\usepackage{cite}
\usepackage{amsmath,amssymb,amsfonts}
\usepackage{algorithmic}
\usepackage{graphicx}
\usepackage{textcomp}
\usepackage{xcolor}
\usepackage{todonotes}
\usepackage{colortbl}
\usepackage{hyperref}

\usepackage{booktabs}

\ifCLASSOPTIONcompsoc
\usepackage[caption=false,font=normalsize,labelfont=sf,textfont=sf]{subfig}
\else
\usepackage[caption=false,font=footnotesize]{subfig}
\fi

\usepackage{enumitem}
\usepackage{upgreek}
\definecolor{azure}{rgb}{0.0, 0.5, 1.0}
\definecolor{cadetgrey}{rgb}{0.57, 0.64, 0.69}
\definecolor{lightblue}{rgb}{0.0, 1.0, 1.0}
\definecolor{lightgray}{rgb}{0.9, 0.9, 0.9}

\newcommand{\orcid}[1]{\href{https://orcid.org/#1}{\includegraphics[height=10pt]{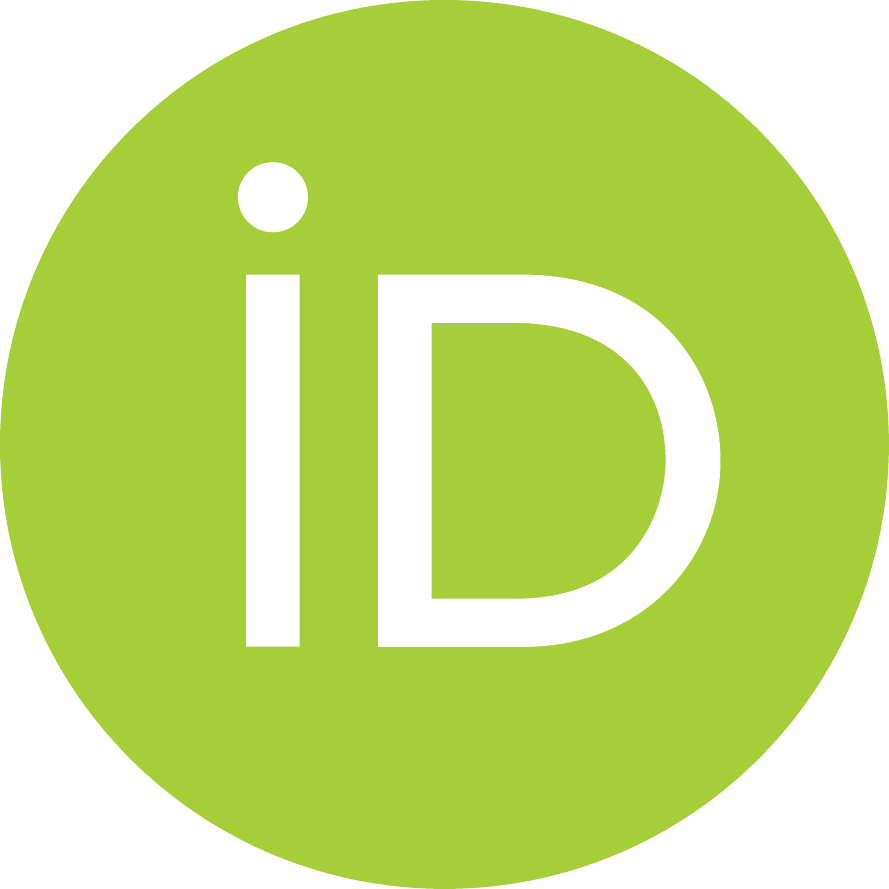}}}

\usepackage{aas_macros}

\usepackage{booktabs}
\def\BibTeX{{\rm B\kern-.05em{\sc i\kern-.025em b}\kern-.08em
    T\kern-.1667em\lower.7ex\hbox{E}\kern-.125emX}}
\begin{document}
\bstctlcite{IEEEexample:BSTcontrol}

\title{From Task-Based GPU Work Aggregation to Stellar Mergers: Turning Fine-Grained CPU Tasks into Portable GPU Kernels
}

\author{
\IEEEauthorblockN{Gregor Dai\ss\orcid{0000-0002-0989-5985}\IEEEauthorrefmark{2}, Patrick Diehl\orcid{0000-0003-3922-8419}\IEEEauthorrefmark{1}\IEEEauthorrefmark{4}, Dominic Marcello\IEEEauthorrefmark{1}, Alireza  Kheirkhahan\IEEEauthorrefmark{1}, Hartmut Kaiser\IEEEauthorrefmark{1}\orcid{0000-0002-8712-2806}, Dirk Pfl\"uger\orcid{0000-0002-4360-0212}\IEEEauthorrefmark{2}
\IEEEauthorblockA{
\IEEEauthorrefmark{1}LSU Center for Computation \& Technology, Louisiana State University,
Baton Rouge, LA, 70803 U.S.A}
\IEEEauthorblockA{\IEEEauthorrefmark{2} IPVS, University of Stuttgart,
Stuttgart, 70174 Stuttgart, Germany\\
Email: Gregor.Daiss@ipvs.uni-stuttgart.de}
\IEEEauthorblockA{\IEEEauthorrefmark{4} Department of Physics and Astronomy, Louisiana State University,
Baton Rouge, LA, 70803 U.S.A. }
}

}

\maketitle

\begin{abstract}
Meeting both scalability and performance portability requirements is a challenge for any HPC application, especially for adaptively refined ones. In Octo-Tiger, an astrophysics application for the simulation of stellar mergers, we approach this with existing solutions: We employ HPX to obtain fine-grained tasks to easily distribute work and finely overlap communication and computation. For the computations themselves, we use Kokkos to turn these tasks into compute kernels capable of running on hardware ranging from a few CPU cores to powerful accelerators. There is a missing link, however: while the fine-grained parallelism exposed by HPX is useful for scalability, it can hinder GPU performance when the tasks become too small to saturate the device, causing low resource utilization. To bridge this gap, we investigate multiple different GPU work aggregation strategies within Octo-Tiger, adding one new strategy, and evaluate the node-level performance impact on recent AMD and NVIDIA GPUs, achieving noticeable speedups.
\end{abstract}


\section{Introduction}
\label{sec:introduction}
Currently, developers of High-Performance-Computing (HPC) applications are faced with a diverse set of supercomputers:
Machines like Fugaku contain a massive amount of compute nodes (158,976), though each node is comparatively weak with just $48$ CPU cores. 
In contrast, machines like Perlmutter have by far fewer (1,536), but much more powerful compute nodes, each containing four NVIDIA\textsuperscript{\textregistered}~A100 GPUs.
This heterogeneity will further be increased by future machines like Frontier (using AMD\textsuperscript{\textregistered}~GPUs) and Aurora (using Intel\textsuperscript{\textregistered}~GPUs), highlighting the importance of both scalability and performance-portability
for any HPC application.

One such application is Octo-Tiger, a C\texttt{++} astrophysics code, used to study stellar mergers \cite{marcello2021octo}.
It is built upon HPX~\cite{Kaiser2020,kaiser_hartmut_2022_6969649}, a distributed asynchronous many-task runtime system (AMT).
HPX allows us to distribute work as fine-grained tasks, easily overlapping computations and communication by defining their dependencies and letting the runtime system handle the concurrency.

However, the very fine-grained tasks that help to scale Octo-Tiger to thousands of nodes cause difficulties regarding performance-portability when running the application on GPU platforms.
On the one hand, the fine-grained tasks (usually suited for one CPU core) allow us to express the maximum amount of parallelism with HPX, to best enable the overlap of computation and communications for widely distributed runs, and to more easily handle small tasks/workloads occurring due to the simulation's adaptive mesh refinement.
On the other hand, to properly utilize GPUs, we need enough work items per GPU kernel to both scale to all compute units (for instance 120 on an MI100 GPU) and to have sufficient resident work items per compute unit to properly hide latencies. In short, we would like to have large enough tasks to turn into GPU kernels that do not starve the device.
Faced with the conflict between both fine-grained and large tasks, it is not enough to merely have compute kernels that are capable of running both on CPUs and GPUs. To achieve optimal performance, it is rather necessary to dynamically adjust the workload per task depending on what device it should run on.

In this work, we investigate and compare three work aggregation strategies to increase the workload per compute kernel within Octo-Tiger, turning fine-grained tasks designed for distributed CPU scenarios into a good match for GPUs.
Firstly, one conventional approach is to simply subdivide the overall discretization of an application into larger sub-problems when building for GPUs than we would for a CPU build, hence increasing the workload per kernel.
In the context of this work, we will refer to this approach as "strategy 1".
Secondly, one can use the ability of GPUs to run multiple independent kernels concurrently, hence relying on the GPU's runtime to implicitly aggregate the kernels on the device side to avoid low resource utilization ("strategy 2"). This approach heavily depends on the abilities of the GPU runtime to handle large amounts of small kernels, as well as the ability of the application itself to launch said kernels with little overhead.
Lastly, instead of relying on the GPU runtime to run independent kernels in parallel as for strategy 2, we can use explicit work aggregation. Thus, we aim to aggregate similar but independent kernels combined on-the-fly into one compute kernel whenever the GPU is starved ("strategy 3").

Past work porting HPX-based codes such as Octo-Tiger to GPU platforms relied on the second strategy:
The GPU implementation of OctoTiger was first introduced in~\cite{daiss2019piz} for Octo-Tiger's gravity solver. It combines multiple NVIDIA\textsuperscript{\textregistered}~CUDA\textsuperscript{\textregistered}~streams together with an HPX-CUDA integration treating CUDA kernels as HPX tasks, achieving good performance.

In turn, in~\cite{diehl2021octo} a similar CUDA implementation of Octo-Tiger's second major module, the hydrodynamics solver, was added. 
It used the same work aggregation strategy as the gravity solver, however, we showed that increasing the work size per kernel launch (by subdividing the grid into larger-than-usual sub-grids, reminiscent of strategy 1) yielded a further node-level speedup for this solver. 
This indicates that the workload per compute kernel is still problematic in the new hydro solver implementation when just using strategy 2. 
Increasing the sub-grid size using strategy 1 further came at the expense of scalability and adaptive refinement, hence motivating us to look for alternative work aggregation strategies beyond these first two strategies.

In this work, we improve upon the state-of-the-art and introduce an implementation for the explicit work aggregation strategy (strategy 3), building on HPX and its accelerator support. This allows us to launch GPU kernels through a special executor that enables the aggregation of kernels into larger kernels on-the-fly.

Additionally, we are moving from CUDA to Kokkos, providing performance portability across different systems~\cite{edwards2014kokkos}.
The existing HPX-Kokkos integration layer that allows us to treat Kokkos kernels as HPX tasks and that allows HPX worker threads to execute Kokkos kernels, is effectively moving away from the Fork-Join model~\cite{daiss2021beyond}. This has already been used for an implementation of the gravity solver. To achieve strategy 3 with the help of both HPX and Kokkos, we extend the hydrodynamics solver by a similar Kokkos implementation in this work. 
For a fair comparison on AMD GPUs we also added a HIP version, by simply reusing the CUDA kernels using the appropriate HIP API calls.

This means that  in this work we compare three different kernel implementations of the hydrodynamics solver: CUDA, HIP, and Kokkos.
For each implementation, we look at results from all three aforementioned GPU work aggregation strategies.
Testing the Octo-Tiger node-level performance on both an A100 and a MI100 GPU, we show that a combination of strategies is vastly superior to the current status-quo of just a single one: We obtain clear speedups on both devices for Octo-Tiger. We further show that the strategies exhibit different performance behavior depending on the GPU vendor, highlighting the need of having alternative strategies at hand for performance-portability if work aggregation is needed.

Overall, this work has three main contributions: 1) The novel on-the-fly work aggregation executor (implementing strategy 3), 2) the implementation of the hydrodynamics module in Kokkos, and 3) a thorough comparison of the new work aggregation strategy with the two existing ones using both the new Kokkos hydro kernels and their previous CUDA counterparts.
For Octo-Tiger itself, our contributions lead to a significant speedup. Beyond Octo-tiger, both the new aggregation executor and the insights gained by comparing different GPU work aggregation strategies can be used to find optimal strategies in other HPX applications.

The remainder of this paper is structured as follows:
Section~\ref{sec:related_work} relates work regarding task-based programming frameworks to GPU support. Section~\ref{sec:hpx} introduces HPX, Section~\ref{sec:octotiger} the scientific application Octo-Tiger. The three work aggregation strategies are introduced in Section~\ref{sec:strategies}, followed by results and their extensive comparison on different systems in Section~\ref{sec:results}. 
\section{Related Work}
\label{sec:related_work}
In this section, we restrict the overview to AMTs with accelerator support, namely CUDA, HIP, and Kokkos~\cite{9485033}. For a more general overview of AMTs, we refer to~\cite{thoman2018taxonomy}. Table~\ref{tab:overview} summarizes the accelerator support. For the AMTs supporting accelerator support, all support NVIDIA GPUs using CUDA~\cite{10.1002/cpe.1631,9150427,SCI:Hol2022b,daiss2019piz,majeti2015heterogeneous,cunningham2011gpu}. The support of AMD GPUs via HIP is provided by HPX solely. Uinath supports AMD GPUs via the Kokkos backend~\cite{SCI:Hol2019a}. In addition to CUDA and HIP, HPX provides Kokkos support~\cite{daiss2021beyond}. Most AMT support acceleration cards nowadays. Now let us have a look into the support of work aggregation. Legion showed aggregation of memory bandwidth of multiple GPUs for Graph Processing~\cite{jia2017distributed}. In addition, a novel dynamic load balancing strategy that is cheap and achieves good load balance across GPUs is presented. For Chapel, a \texttt{GPUIterator}~\cite{hayashi2019gpuiterator}, which supports hybrid execution of parallel loops across CPUs and GPUs, is available. 
However, these solutions are unlike our work, as we use a bottom-up approach, aggregating small HPX tasks on-the-fly into larger GPU kernels.

\begin{table*}[tb]
    \centering
    \caption{Accelerator support for various AMTs. We restricted ourselves to AMTs with support for NVIDIA and AMD GPUs.}
    \begin{tabular}{l|lllllllll}
         & HPX~\cite{Kaiser2020} &  Chapel~\cite{chamberlain2007parallel} & Charm\texttt{++} &  Legion~\cite{bauer2012legion} & Uintah~\cite{germain2000uintah} & ParSec~\cite{bosilca2013parsec} & StarPU~\cite{AugThiNamWac11CCPE} &  X10~\cite{ebcioglu2004x10} & UPC\texttt{++}~\cite{zheng2014upc++}  \\\midrule
        CUDA & \checkmark~\cite{daiss2019piz} & \checkmark~\cite{9150427} & \checkmark~\cite{lifflander2012dynamic} & \checkmark & \checkmark~\cite{SCI:Hol2022b} & \checkmark~\cite{bosilca2011performance} &\checkmark~\cite{10.1002/cpe.1631} & \checkmark~\cite{majeti2015heterogeneous} & \checkmark \\
        HIP & \checkmark~\cite{daiss2021beyond} & & & & \checkmark & & & & \checkmark  \\
        Kokkos & \checkmark~\cite{daiss2021beyond} & & & & \checkmark~\cite{SCI:Hol2019a} & & & &  \\
    \end{tabular}
    \label{tab:overview}
\end{table*}


\section{C\texttt{++} standard library for parallelism and concurrency}
\label{sec:hpx}
One asynchronous many-task system runtime system with distributed capabilities is the C\texttt{++} standard library for parallelism and concurrency, HPX~\cite{Kaiser2020}. One major difference of HPX from other AMTs is that HPX's API is fully conforming with the recent and upcoming C\texttt{++} standard~\cite{cxx14_standard,cxx17_standard,cxx20_standard,cxx23_standard}. Note that other AMTs are written in the C\texttt{++} programming language, but HPX's API follows the definition of the C\texttt{++} standard for the asynchronous programming and the parallel algorithms. We refer to the references~\cite{hpx_pgas_2014,Kaiser:2015:HPL:2832241.2832244,Heller2016,Kaiser2020} for more details about HPX. In this paper, we use HPX for the following two purposes: \textit{1)} the coordination of the synchronous execution of a multitude of heterogeneous tasks (both on CPUs and GPUs), thus managing local and distributed parallelism while observing all necessary data dependencies; and \textit{2)}  as the parallelization infrastructure for launching HIP/CUDA-kernels on the GPUs via the asynchronous HPX backend.


\section{Octo-Tiger}
\label{sec:octotiger}
As a prototypical adaptive mesh refinement (AMR) code with non-trivial physics we consider Octo-Tiger, an astrophysics code modelling stellar mergers~\cite{marcello2021octo}. Octo-Tiger uses a fast-multipole method (FMM) to solve for gravity~\cite{2017AJ....154...92M}. The implemented FMM globally conserves both linear and angular momenta up to machine precision. To model and discretize the hydrodynamics components, a finite volume method using AMR is employed. Coupling the FMM with the hydro solver allows global conservation of energy and linear momentum up to machine precision, a major strength of Octo-Tiger. 


\subsection{Scientific Application and Previous Results}

Octo-Tiger is designed to model interacting binary star systems. A binary star system consists of two stars, bound to one another by gravity. When they are close enough together, they interact by exchanging mass. Sometimes this mass transfer is stable and long-lived over millions of years. Sometimes it is unstable, leading to a catastrophic disruption of one of the binary's components. When this happens and if the system is massive enough, a Type Ia supernova results. Less massive systems result in the merger of the disrupted star with its companion, leading to the formation of another star. The helium rich R Coronae Borealis stars are thought to originate from a merger of two white dwarfs.

Octo-Tiger models such systems as self-gravitating fluids, governed by the laws of hydrodynamics and Newtonian gravity. The code has been used to investigate the origins of R Coronae Borealis stars (\emph{e.g.}\ \cite{2018ApJ...862...74S, 2019MNRAS.488..438L}), the merger of bipolytropic stars~\cite{2018MNRAS.481.3683K}, and the possibility that the star Betelgeuse is the outcome of a merger~\cite{2020ApJ...896...50C}. Presently, Octo-Tiger is used for the investigation of merging double white dwarfs as well as the merger of a contact binary, V1309 Sco.

\subsection{Hydro Solver}
Octo-Tiger solves the inviscid Euler equations. This set of hyperbolic differential equations governs the conservation of mass, momentum, and energy as a fluid evolves. Octo-Tiger is a grid based code, using Cartesian adaptive mesh refinement to discretize the fluid variables. Octo-Tiger uses the piecewise-parabolic method~\cite{COLELLA1984174} to compute the values of the evolved variables at 26 quadrature points on the surface of each computational cell - one for the centers of each cell face and each cell edge, and one for each cell vertex. With the reconstructed variables, the fluxes are computed at these points using the central upwind method as described by \cite{kurganov}. They are integrated using Newton-Cotes quadrature to obtain the total flux through a cell face.
The maximum allowed time-step size is related to the ``Courant condition": The time-step size has to be at most the minimum time it takes a signal to cross a computational cell's width. Exceeding this time-step size results in errors in the solution that grow rapidly with time.  
If we double the resolution of the model without altering the model's size, this signal crossing time will be roughly cut in half, reducing the allowed time-step size by the same factor.

The AMR feature of Octo-Tiger is designed to refine around interesting areas of the binary. One level of refinement is assigned to each component as a whole, allowing a smaller component to be modelled with higher resolution than its companion. The cores of stars with core/envelope structures can be given additional levels of refinement. 
Recently, the use of gradient-based refinement is being investigated as well to additionally refine the star's atmospheres.

\subsection{Previous Scalability/Performance}
Performance on up to $5400$ GPUs and $64,800$ cores on  CSCS's Piz Daint was shown in~\cite{daiss2019piz}. Performance on up to $658,784$ Intel Knight’s Landing cores with a parallel efficiency of $96.8\%$ using billions of asynchronous tasks was demonstrated in~\cite{heller2019harnessing} on NERSC's Cori. Performance on ORNL's Summit was shown in~\cite{diehl2021octo}. 


\begin{figure}
    \centering
    \subfloat[\label{fig:strategy:1}]{
    \includegraphics[width=.30\linewidth]{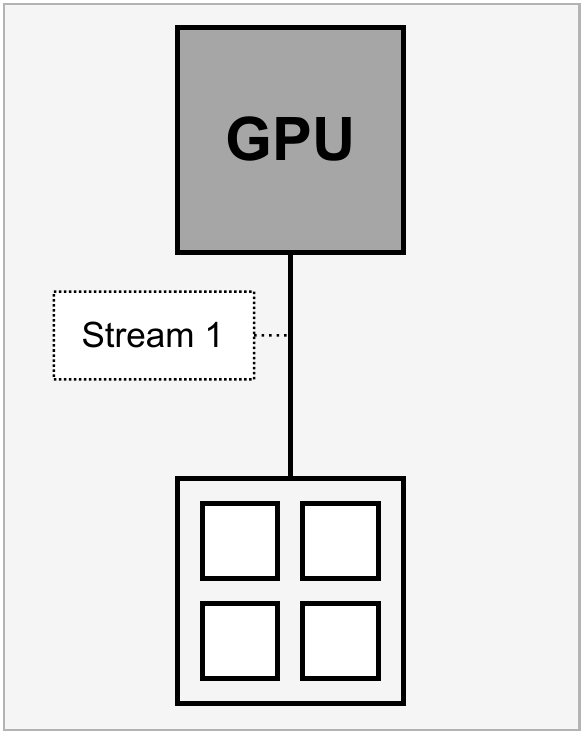}  
    }
    \subfloat[\label{fig:strategy:2}]{

    \includegraphics[width=.300\linewidth]{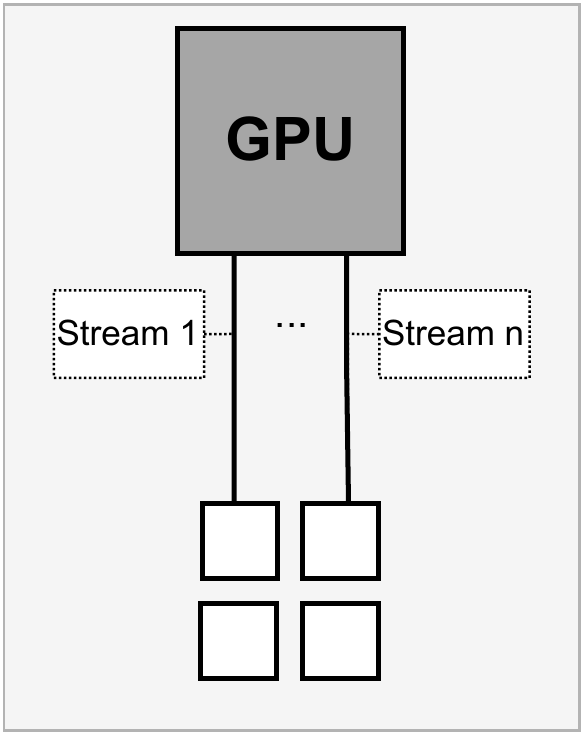}  
    }
    \subfloat[\label{fig:strategy:3}]{
    \includegraphics[width=.30\linewidth]{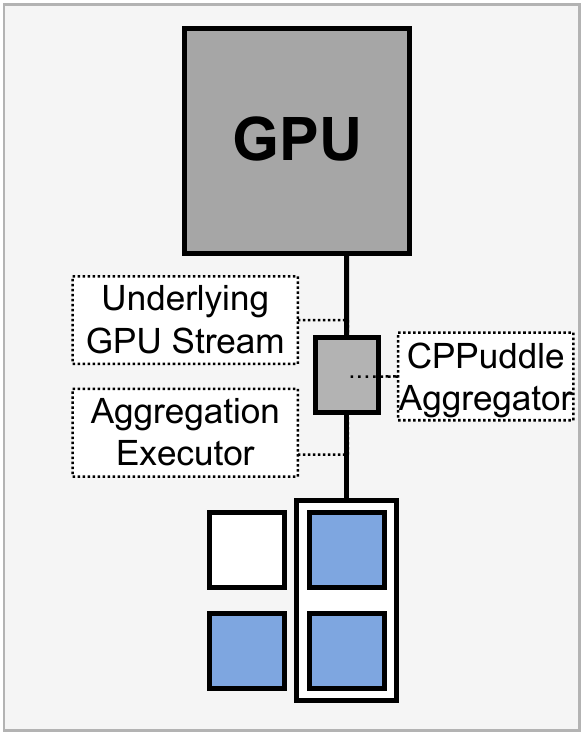}  
    }
    \caption{Aggregation strategies: (a) Larger sub-problems: Increasing sub-grid size, (b) Implicit work aggregation: Interleaving independent GPU kernels by using multiple GPU executors (streams), and (c) Explicit work aggregation: Marking compatible tasks (blue) which might be aggregated together to a larger kernel if the hardware is currently busy.}
    \label{fig:my_label}
\end{figure}

\section{Aggregation Strategies and GPU Implementation details}
\label{sec:strategies}
In this section, we first give some details about the implementation of the hydro solver, especially regarding the workload per GPU kernel.
As these numbers motivate our need for larger numbers of GPU work items, we continue with the introduction of three strategies for increasing the size of the GPU kernels.
For each of the strategies, we first introduce the high-level idea, mention the strategy's requirements, then briefly talk about the implementation details and their respective benefits and challenges.
While the first two strategies have already been used with Octo-Tiger, the last strategy (strategy 3) is a novel addition of this work.
\subsection{The Need for Work Aggregation in Octo-Tiger}
Octo-Tiger uses an adaptive octree as its data-structure, with each node consisting of a whole sub-grid for efficiency. All required variables for the hydro solver are located on such sub-grids.
The (uniform) size of those sub-grids is determined at compilation time.
By default, Octo-Tiger uses a $8^3$ sub-grid size, however, Octo-Tiger can be configured to use larger sub-grids as well.
The relevant work of the hydro solver is largely done by two GPU kernels that are invoked multiple times for each sub-grid in each time-step: The \texttt{Reconstruct} kernel and the \texttt{Flux} kernel.
Together with smaller auxiliary methods (both on CPU and GPU), these two kernels implement the hydro solver.
Compared to the aforementioned auxiliary methods, the \texttt{Flux} and \texttt{Reconstruct} kernels constitute the vast majority of the runtime.
Each of these kernels were originally developed to be executed by one CPU core in the form of an HPX task (achieving multicore usage by running those HPX tasks concurrently for different sub-grids).
Since then, the hydro solver was ported to GPU.
We added and benchmarked a CUDA version in~\cite{diehl2021octo}, yielding a significant speedup over the previous parallel CPU implementation.
However, we also found further performance improvements when increasing the sub-grid sizes (while using fewer sub-grids to maintain the same overall number of grid cells).

For this work, we now further extend the GPU support in the hydro solver by adding both a HIP version and a Kokkos version\footnote{Added mostly in https://github.com/STEllAR-GROUP/octotiger/pull/389 and the latter part of https://github.com/STEllAR-GROUP/octotiger/pull/365}. The HIP version shares most parts of its code with the CUDA kernels but needed adaptation in how the kernels are invoked -- overall it should perform similarly to its CUDA counterpart.
We add a Kokkos implementation for each kernel of the hydro solver, as porting Octo-Tiger's gravity solver to Kokkos already yielded promising results in \cite{daiss2021beyond}.
These hydro Kokkos kernels are a straightforward port of their respective CUDA versions (for example, mapping a shared memory implementation to an implementation using Kokkos \texttt{Team Policies} and \texttt{Scratch Memory}). They also use the same launch configurations (number of blocks, work items per block) as their CUDA counterparts.

Thus, while the \texttt{Reconstruct} and \texttt{Flux} kernels have overall changed a lot, going from single-core to GPU kernel, their workload has not changed:
As the sub-grids are meant to be distributed across multiple compute nodes, the kernels themselves still operate on only one sub-grid at a time.
While this gives us an easy way to distribute the workload across the compute nodes by simply distributing the sub-grids, it also leads to tiny GPU kernels in terms of work items.

To be specific, given the default $8^3$ sub-grid size, the kernels process inputs of up to $14^3$ (as we have a ghost layer with a thickness of $3$) and both have $10^3$ work items (as we need to reconstruct values and calculate the flux for the innermost ghost layer as well).
As mentioned, in the original design each of those kernels was meant to be executed by one HPX task (and hence run on $1$ CPU core each). For CPUs, the amount of work is sufficient to ensure efficient parallelization.
However, this does not translate well to GPUs, where more work items enable us to hide latencies better by executing with a higher occupancy per compute unit (called CU for AMD devices, SM for NVIDIA ones) and further enable the scaling of execution to all CUs available in the GPU by launching enough blocks.

For example, we mentioned that the \texttt{Reconstruct} kernel operates on  $10^3$ cells given the default sub-grid size.
To do so, the kernel launches 8 blocks, each with $128$ work items. Similarly, the Flux kernel works on $10^3$ cells. However, it can further parallelize over the dimension. Given that Octo-Tiger is 3D, and we still use $128$ work items per block, we therefore have $24$ blocks per \texttt{Flux} kernel invocation.

This means neither the \texttt{Reconstruct}, nor the \texttt{Flux} kernel, launch enough blocks to even scale to the $120$ CUs of an AMD MI100 GPU, or the $108$ SMs of an NVIDIA A100 GPU, let alone provide enough work items to efficiently hide latencies with only one block resident per CU. This will issue only become more pronounced for the next generation of GPUs, as an AMD MI250 contains $208$ CUs. 
For reference, the runtime of just one such $8^3$ \texttt{Reconstruct} CUDA kernel is roughly 300$\upmu$s on an A100  (150$\upmu$s for a \texttt{Flux} kernel).

Hence, we need to aggregate these fine-grained CPU tasks into larger GPU tasks to efficiently use the GPUs.
We can rely on the fact, that while our kernels are small, they are executed for each sub-grid, meaning we have a high number of independent kernel invocations in every solver iteration (production-level scenarios with Octo-Tiger can have hundreds of thousands of sub-grids, distributed across hundreds of compute nodes).
To this end, we show three strategies to aggregate work, or alternatively avoid splitting it to begin with, to increase the workload per GPU kernel, thus leading to more efficient GPU usage.

\subsection{Strategy 1: Varying Sub-Grid Sizes (Larger Sub-Problems)}
\textbf{The high-level idea:}
There is no easier way of increasing the workload per GPU kernel than by simply dividing the computational space into larger sub-problems than usual (thus avoiding the small workload per GPU kernel invocation to begin with), see Figure~\ref{fig:strategy:1}.

\textbf{Requirements:} 
This requires the application to be written in a way that supports adjusting the size of said sub-problems without changing the size of the overall problem (or changing the output).

\textbf{Implementation details:}
In Octo-Tiger, we use sub-grids to subdivide our overall grid into smaller problems. This is how Octo-Tiger distributes work over multiple compute nodes in distributed runs. Each node works on a number of sub-grids that can communicate with other sub-grids via HPX parcels (messages).
As we can configure the sub-grid size in Octo-Tiger during compile-time, we can directly influence the number of work items per GPU kernel.
For example, moving to $16^3$ sub-grids, increases the number of work items for the aforementioned GPU kernels to $18^3$.

\textbf{Benefits / Challenges:}
This is the most straightforward way of increasing the workload per kernel invocation, in case the application supports it.
In Octo-Tiger, this strategy has a second advantage: The grid is distributed across a smaller number of sub-grids.
As each sub-grid has its own ghost layer, this reduces the overall number of ghost cells.

However, the strategy has several disadvantages, making its use problematic for Octo-Tiger: 

\begin{enumerate}[itemsep=1pt, topsep=1pt, partopsep=0pt]
    \item Larger sub-grids result in more refinement outside the actual area of interest, resulting in computations that are not needed.
    \item Larger sub-grids lead to a lesser number of sub-grids to distribute onto compute nodes, thus hitting scalability limits sooner for distributed scenarios. For Octo-Tiger this was shown in~\cite{diehl2021octo} where the scalability in distributed runs became worse for larger sub-grids, even though the individual kernels ran more efficiently.
    \item Increasing the size of sub-grids has implications on other parts of the code. For instance, the FMM gravity solver works best with smaller sub-grids and more levels of refinement as it uses the tree structure to avoid computations through approximation.
\end{enumerate}
Of course, whether each (or any) of these disadvantages applies to other codes depends on the application in question.
Overall, this strategy is highly application-specific.
\subsection{Strategy 2: Interleaving Independent GPU Kernels as Tasks (Implicit Work Aggregation)}

\textbf{The high-level idea:} 
Instead of worrying about the workload of the individual GPU kernels, we can take advantage of the large number of overall kernel launches and run them concurrently on the GPU, allowing the GPU runtime to aggregate the kernels for us on the device side (hence implicit work aggregation), see Figure~\ref{fig:strategy:2}.

\textbf{Requirements:}
Here, each CPU thread needs to handle multiple kernel launches simultaneously, as we have more kernels to launch than CPU threads (and more than the maximum number of $128$ concurrent GPU streams). This means that we need to avoid any overhead when launching the GPU kernels. Each launch needs to be completely non-blocking, and the CPU threads should never suspend while waiting for GPU results of a kernel as long as there are more kernels to launch (as each needs to work on launching more kernels).

\textbf{Implementation details:}
To fulfill these requirements, we leverage features of the task-based runtime system.
HPX itself works well together with CUDA and HIP, as it is able to treat CPU-GPU data transfers and GPU kernels as HPX tasks, thus it is able to tightly integrate GPU-related work with other tasks such as inter-node communication or auxiliary CPU tasks.
Building on this idea we enabled the integration of Kokkos kernels into the HPX task graph similarly~\cite{daiss2021beyond}.

We can utilize this HPX-CUDA/HIP/Kokkos integration in order to run multiple independent GPU kernels concurrently on the GPU.
This has been possible for a long time, for example using CUDA streams.
However, given the short-running nature of our kernels, each of our CPU threads needs to handle multiple, ongoing, asynchronous GPU kernels and keep track of their eventual completion to use the results for various new tasks.
This is made easy by using HPX:
Since we can treat kernels as HPX tasks, the HPX runtime system keeps track of them for us, freeing us from manually handling the synchronization of each GPU kernel invocation.
Each kernel invocation has an attached HPX future that becomes ready once the kernel has finished running, meaning a thread can simply launch a kernel in an HPX task, while immediately continuing to work on another task, and return to the original task once the GPU kernel is complete to use its results.
This comes with the advantage that no CPU thread is ever blocked by actively waiting for a GPU kernel to finish, as long as there are any HPX tasks left it can work on instead.

Further, we need to keep the overhead of launching kernels as low as possible.
Particularly, we cannot afford to have memory allocations ({\it mallocs}) on the GPU or the creation of the GPU executors, as the mallocs (or the creation of GPU stream in case of the executor) cause the entire device to synchronize!
However, pre-allocating all required pinned host buffers and GPU buffers (including temporary and ghost cell buffers) would increase the memory requirements in Octo-Tiger.
Hence, we prefer being able to allocate temporary GPU buffers at low overhead.

To this end, we use a dynamic pool of buffers that is provided by \texttt{CPPuddle}, a small utility library for task-based GPU programming. If a buffer of the requested type and size does not exist, it will be created (causing a Malloc). However, once the current task is done with it, the buffer will not be freed yet, but instead goes back into the pool to be recycled by another task (thus avoiding any Mallocs).
This scheme works particularly well since, given the task-based structure, we launch the same tasks potentially thousands of times for different sub-grids in each time-step.
We can configure the type of memory used by the pool by passing it an underlying allocator type, making it easy to use it either for HIP/CUDA device memory or pinned host memory buffers.
In turn, the pool can be accessed using the allocators provided by \texttt{CPPuddle} itself. These make it easy to integrate the pool with normal vectors, or even Kokkos Views.

Lastly, to prevent the creation of temporary device executors (and thus the creation of their underlying GPU streams that would synchronize the entire device) we use a pre-allocated pool of executors that is initialized during the start of the application.
This is again managed by \texttt{CPPuddle}. If the application needs a GPU executor to access a task, it can extract one from the pre-allocated pool that provides either a round-robin or a load-balancing scheduling across the executors.

This strategy was first considered in~\cite{daiss2018octo}, then fully implemented in~\cite{daiss2019piz} for Octo-Tiger's gravity solver and HPX+CUDA where it achieved good GPU performance but was still Octo-Tiger specific.
The approach was later refined (and generalized) in~\cite{daiss2021beyond} using \texttt{CPPuddle} and Kokkos. 

\textbf{Benefits / Challenges:} 
Combining this low-overhead way of accessing GPU buffers/executors to launch GPU kernels, together with the ability of HPX to easily track multiple kernels launched without blocking the CPU thread, we are able to launch GPU kernels quickly enough to have them run concurrently despite their short, individual runtime. Hence, we are able to utilize the GPU effectively despite the small and short running GPU kernels. 
As a secondary advantage, this automatically enables us to interleave the GPU kernels with any other tasks, such as CPU-GPU data-transfers, CPU tasks, and internode communication as dictated by the dependencies in the HPX Task graph.
Unlike strategy 1, this strategy is also application-independent, it can work in any HPX application by using the respective GPU executors (and if necessary, the allocators to access the \texttt{CPPuddle} memory pools).

However, it is also clear that this technique has its limitations, dictated by the GPU runtime system and related API overheads.
If the kernels are becoming too small, we may not be able to run a sufficient number concurrently to fully utilize the GPU.
Furthermore, this technique also causes a large amount of CUDA/HIP API calls to be executed, thus adding more overheads.
Bottlenecks in the API that are unnoticeable in other applications with fewer API calls can be the limiting factor when employing this strategy.
Thus, this strategy might perform differently depending on the employed GPU architecture.

\subsection{Strategy 3: Buffering Kernel Launches (Explicit Work Aggregation)}
\textbf{The high-level idea:}
Using strategy 2, kernel launches on the same executor are just queued up in its CUDA (or HIP) stream, even if said executor is currently already busy running another kernel.
A different approach is to combine multiple of those kernel invocations on-the-fly and launch them as one larger kernel.
This reduces the load on the API (as there are fewer overall kernel launches), the contention between shared GPU resources (pre-allocated buffers and executors in \texttt{CPPuddle}),
and it increases the size of the kernels running on the GPU, making us less dependent on the GPU runtime running separate kernels in parallel as in strategy 2.

To achieve this, we rely on the fact that we have many tasks doing the same thing on different sub-grids.
Essentially, we want to be able to define tasks within our task graph that can be executed together by a team of tasks, ideally in a way that is indistinguishable from a normal kernel launch, except by using different executors/allocators.
Overall, this is very much in the spirit of HPX, as we still define fine-grained tasks, but give hints which ones can be aggregated for more efficient GPU kernel launches if the underlying hardware is currently busy. See Figure~\ref{fig:strategy:3}.

\textbf{Requirements:} We have multiple requirements for this:
We want to aggregate the tasks launching specific GPU kernels on-the-fly in case the currently used GPU executor is already busy.
We want this to be non-blocking on the CPU side by leveraging HPX futures.
Furthermore, we want this to be minimally invasive in the application code.

As GPU kernels are often not just launched in a single line, but further contain some staging area where buffers for the GPU are filled, we want to be able to define an entire region of code that is allowed to be aggregated with other tasks executing the same code region for a different sub-grid.
Besides defining a code region that is allowed to be aggregated (aggregation region) and using a special set of allocators and executors, aggregated kernel launches should not be different from the normal non-aggregated ones.

When an HPX task hits such an aggregation region, it will only continue if there is a non-busy GPU executor. Otherwise, the task will wait until the executor becomes free and then enter the code region, potentially together with other tasks that tried to execute the same code region with a different sub-problem (or sub-grid in the case of Octo-Tiger). Note again, that the CPU thread is not blocked by the HPX task waiting, as it simply continues to execute other HPX tasks until the current one becomes ready.
In the case where multiple tasks enter the region together, as they got aggregated while waiting, they will share the same executor and the same allocators, which enables the aggregation under the hood. Buffers allocated with the allocators are part of a consecutive chunk of memory that is shared between the tasks, functions (like kernel launches and GPU data transfers) are only executed when all participating tasks have called them. This allows the tasks to work together in filling all staging-area buffers (each on their own chunk) for the GPU kernel launch (thus sharing the burden of communication, pre-processing and data copies). The GPU launch itself is just a larger kernel executing the same function for all chunks.

To keep the implementation of this strategy simple, we define one constraint:
Within such an aggregation region, we have a "Single-GPU-workload-Multiple-Tasks" semantic.
This means either all participating tasks launch a kernel (or allocate an aggregated buffer) or none do. Further, it means the order of allocations and kernel launches is the same on all tasks within the defined aggregation region.
However, this only holds true for functions launched via the executor and buffers allocated via the allocators (in practice only for the GPU API calls and buffers). The tasks can operate independently otherwise, doing communication and pre- / post-processing to fill their part of the aggregated buffers.

\textbf{Implementation details:} We implemented\footnote{Added in https://github.com/SC-SGS/CPPuddle/pull/12} this approach in \texttt{CPPuddle} to have it available independent of Octo-Tiger (as its executors/allocators themselves are independent of the application).
It includes a separate Executor Pool (which will be created for each aggregation region defined in the application), as well as the aggregation executors and allocators used by the tasks to facilitate aggregated kernel launches.

In a nutshell, \texttt{CPPuddle} leverages HPX to keep CPU threads from waiting once they hit an aggregation region, and further uses atomic counters and mutexes for concurrency control to coordinate the shared kernel launches and buffer allocations under the hood.
Within an aggregation region, the first task that encounters a shared buffer allocation (\emph{i.e.}\ an allocation done with the aggregation allocator), allocates the actual buffers (or reuses a currently unused buffer of the correct size from one of the \texttt{CPPuddle} buffer pools), the other tasks hitting the same allocation will use the same buffer but different chunks of it. Since all tasks are required to have the allocations in the same order, this can be accomplished with a simple atomic counter.

Similarly, the last task encountering a kernel launch will actually launch the aggregated kernel. The previous tasks encountering the same kernel launch simply signify that the other tasks are done filling their respective parts of the input buffers for the task.
Internally, the aggregation executors are connected to a parent executor, which is not exposed to the user. All necessary communication between the executors is done over this parent. To avoid a bottleneck here, we can have an arbitrary number of parents.

The parent executor in turn has an underlying GPU executor which is used to launch the aggregated kernels and a direct interface to \texttt{CPPuddle} to request pre-allocated buffers that are currently not in use. Depending on the number of GPU executors used, parents might share one GPU executor, which causes them to simply queue launches into the same CUDA/HIP stream (making it more likely the GPU executor is busy when tasks hit an aggregation area).
By using multiple parents with multiple (distinct) underlying GPU executors, this also effectively allows us to combine this strategy 3 with the previously mentioned strategy 2 where we relied on concurrent execution with multiple GPU executors.

Lastly, we define a parameter which defines the maximal number of tasks that are allowed to enter an aggregation region together. This acts as an optional, second criteria to enter an aggregation region. If enough tasks are waiting together (thus reaching the defined maximum), they will enter the region whether the underlying GPU executor is ready or not. This prevents the system from aggregating too large kernels, which would not yield any more runtime benefits but might instead cause unnecessary overhead as too many tasks are waiting to enter aggregation regions. This will also be used in Section~\ref{sec:results} to control how many kernels can be aggregated together, enabling us to test different configurations.

\textbf{Benefits / Challenges:} 
As mentioned above, this approach has distinct benefits over strategy 2 where we relied solely on concurrent execution of smaller GPU kernels: Increased kernel size (making it easier to completely utilize the GPU without relying on the API to run kernels concurrently), reduced load on the GPU runtime (fewer API calls), reduced load on shared GPU buffers/executors.

Compared to a more static approach (like defining specific sub-grids to always launch an aggregated GPU kernel), the dynamic nature of this approach also has some benefits:
Since it is decided on-the-fly for each sub-grid (based on the load of the GPU) if it is part of an aggregated GPU kernel launch or not, it is easy to add, remove or migrate sub-grids over the course of the simulation, which is essential in an adaptive, long-running, distributed simulation such as Octo-Tiger, where the grid is newly refined and rebalanced every few time-steps.

However, as with the other strategies, there are downsides again as well.
Firstly, on its own, this approach does not interleave the CPU-GPU data transfers like strategy 2 does. However, that can be overcome by using multiple parent executors again, in turn using multiple underlying GPU executor (effectively combining this approach with strategy 2).
The other downside is simply due to it still low-level proof of concept implementation:
In addition to defining the aggregation regions in the application code (which is essentially just a matter of passing a C\texttt{++} lambda to a function and give it a name), GPU kernels also need to understand the additional index (identifying the task which launched this part of the kernel) to map the arrays correctly.
This is easier to use with the Kokkos kernels as we have implemented a method automatically returning the correct subviews given said index.

Overall, the current proof-of-concept implementation\footnote{Integrated into Octo-Tiger's hydro solver within the pull request https://github.com/STEllAR-GROUP/octotiger/pull/426} already enables us to take a closer look at the performance, allowing us to gauge the worth of the approach before refining it further.
\section{Results}
\label{sec:results}
In this section we look at the performance impact of the various aggregation strategies introduced in the previous section.
To this end, we first introduce the utilized scenario, then explain what parameters are being varied for the work aggregation benchmarks. Afterwards, we move on to discuss the results.
\subsection{Scenario and Hardware}

As we integrated the third work aggregation strategy in Octo-Tiger's hydrodynamics solver, we limit ourselves to a pure hydrodynamic scenario in this work: The Sedov–Taylor Blast Wave scenario \cite{1946JApMM..10..241S}.
This scenario was previously used to benchmark Octo-Tiger's hydrodynamic module and to verify its output, as the scenario has an analytical solution in 3 dimensions \cite{marcello2021octo}. 

For the comparison purposes of this paper, we change the grid structure for this scenario by turning off the AMR. This causes the scenario to have the same number of overall cells when using an octree with three levels and $8^3$ sub-grids compared to one that has two levels but uses $16^3$ sub-grids.
The exact scenario size in terms of the grid cells can be found in Table~\ref{results:scenario-size}. This table also includes the overall number of GPU kernel launches per time-step. This number is the result of having 5 kernels that are launched per sub-grid per hydro-solver iteration, with each time-step including three iterations. As long as kernels are not aggregated together by strategy 3, we thus have separate $7680$ GPU kernel calls and $15360$ CPU-GPU data-transfers per time-step. The majority of the required runtime is caused by two of those five kernel types launched per sub-grid, namely the \texttt{Flux} kernel and the \texttt{Reconstruct} kernel. However, as the other kernel launches contribute to overall GPU overhead, we include them here as well.
The timings are measured by running Octo-Tiger for $15$ time-steps and then dividing the overall computation time 
by the number of time-steps to get the average.
All floating point calculations are in double precision. 
  
We use two machines for running the work aggregation benchmarks with this scenario: 
First, an NVIDIA node, containing an NVIDIA A100 GPU, and Intel\textsuperscript{\textregistered}~Xeon\textsuperscript{\textregistered} Platinum 8358 CPU.
Second, an AMD node, containing an AMD MI100 GPU, and an AMD EPYC\textsuperscript{\texttrademark} 7H12 CPU.
We use $32$ CPU cores on both machines to keep the results more comparable between the machines, even though the AMD node would have $32$ more cores in its socket. 
Using both an NVIDIA- and an AMD GPU allows us to test all three work aggregation strategies using all major kernel implementations (CUDA/HIP, and Kokkos using either the CUDA or HIP execution spaces respectively).

Software-wise, we use the following versions (git commits) of Octo-Tiger (\texttt{8a895d8f}), HPX (\texttt{04824da3e8}), Kokkos (\texttt{ba0caeeb1}), HPX-Kokkos (\texttt{20a4496}) and \texttt{CPPuddle} (\texttt{8507da5}). We use clang 12 with CUDA 11.6, and clang 13 with ROCm 4.5.2.

\subsection{Work Aggregation Parameters}
For work aggregation strategy 1, we vary the sub-grid size between $8^3$ (Octo-Tiger's default) and $16^3$. 
As mentioned, to keep the overall number of cells the same, we turn off the AMR and change the maximum tree level according to the sub-grid size used. 
This makes the results with the different sub-grid sizes more comparable, as they solve the same scenario (\emph{i.e.}\ have the same number of cells and thus the same output). 

However, computation-wise there are still differences of course, namely the reduced number of ghost cells when using $16^3$ sub-grids. This leads to more copy operations when using $8^3$ sub-grids, as well as more  overall computations and memory accesses as both the \texttt{Flux}- and the \texttt{Reconstruct} kernels have to access the entire sub-grid and the ghost cells, as well as calculating results for the innermost ghost layer (thickness 1) for the subsequent post-processing that obtains the final results for a time-step. We still include the runtimes for $16^3$ as it helps to put the results of the other 2 strategies into perspective, however, the optimal runtime (using multiple strategies) with $16^3$ should be viewed as lower bound for the best runtimes we can get out of $8^3$ sub-grids.

To evaluate strategy 2, we increase the number of GPU executors that are used.
These are only allocated once at the beginning of the simulation (by using a \texttt{CPPuddle} executor pool) and henceforth shared between all cores (using a round-robin scheduling) to facilitate overlapped GPU kernel launches and data transfers.
Setting this parameter to one (and strategy 3 disabled) effectively disables concurrent GPU kernels, as well as disabling the overlapping of GPU kernel execution with CPU-GPU data transfers.
Setting the parameter to zero turns off GPU execution entirely and only allows for kernels to be executed on the CPU.

Lastly, for strategy 3, we vary the maximum number of kernels that can be aggregated into one large kernel. 
As the aggregation stops as soon as the underlying GPU stream becomes idle, the system can (and will) aggregate fewer kernels than this maximum number, especially when this work aggregation strategy is combined with other strategies.
Setting this parameter to $1$ turns off this work aggregation strategy.

Overall we have three parameters to vary: The sub-grid size, the number of GPU executors and the maximum number of aggregated kernels.

\subsection{Benchmark results}
The results of the work aggregation tests using the Sedov Blast Wave scenario can be found in Table~\ref{results:times}.
They can be categorized into three parts: 
Initial runs using only the CPU for comparison, the three GPU work aggregation strategies on their own, and lastly the best combinations of the strategies that we found.

The CPU-only results provide a baseline for the later runs of the GPU-accelerated version.
Furthermore, the aforementioned differences in the workload between different sub-grid sizes become more visible here:
Single Core runtime is reduced when using $16^3$ sub-grids (due to reduced number of ghost cells), however, the parallel efficiency worse as we have fewer sub-grids to distribute over the cores (a sub-grid is only worked on by one core at a time).

Looking at the first GPU-accelerated run with $8^3$ sub-grids and only $1$ GPU executor (thus, without any work aggregation), we can actually see a slowdown compared to CPU-only runs.
It is worth noting that even this is already a version where all GPU kernel launches are completely asynchronous and non-blocking (using the HPX-CUDA integration), so this slowdown is not due to the host CPU threads being blocked, but can instead be attributed to the GPU being starved by only executing small kernel launches from $1$ GPU executor.

Increasing the work size for each of those kernel launches by moving to a larger sub-grid size does indeed alleviate this issue as expected: We get better overall runtimes per time-step with sub-grids of $16^3$.
Comparing to the run with the smaller sub-grids, we are reducing the complete runtime per time-step from $869$ms to $159$ms with CUDA, and similarly going from $1404$ms to $224$ms using the HIP kernels. 

Looking at the performance of the Kokkos kernels, it is notable that while they perform very similarly to the CUDA kernels on the NVIDIA hardware, on the AMD hardware things look different.
Here, the Kokkos kernels perform noticeably worse than the HIP kernels. This might be at least partly due to the way Kokkos launches kernels with HIP, as it internally uses a pre-allocated array of buffers here to manage the concurrent kernel launches. Once a HIP execution space exceeds a certain number of Kokkos kernel launches it wraps around and starts reusing the pre-allocated buffers from the beginning. To ensure all kernels are done with the buffers, it synchronizes the execution space, meaning it blocks the current host thread until any remaining, still running kernels finish. Worse, this also blocks all other threads from launching kernels on this execution space, as the access to the array is protected with a mutex that the currently blocked thread holds. However, this performance difference on the AMD hardware needs to be further investigated in future work for a more complete picture.

Moving to the results of strategy 2, increasing the number of GPU executors works well on NVIDIA hardware for both the CUDA and the Kokkos kernels.
Here, we achieve runtimes of $132$ms for the CUDA run, $146$ms for the Kokkos run, thus achieving a clear speedup over the $32$-core CPU-only run and even over the previous GPU run with $16^3$ sub-grids (using strategy 1).
However, the HIP runtime seems to struggle more with the concurrent kernel launches using multiple executors:
On HIP (and Kokkos using the HIP backend) we still see a speedup, however less so. Using HIP, we achieve a runtime of $584$ms ($774$ms for Kokkos), which is far slower than the runs with $16^3$ sub-grids.
Indeed, even using the maximum number of GPU executors, we still struggle to match the performance of the ($32$-core) CPU-only run ($518$ms).

To evaluate the performance impact of strategy 3, we are back to 1 underlying GPU executor, but we start aggregating kernels together in case said underlying executor is busy.
Strategy 3 already works well on the NVIDIA node ($156$ms CUDA, $165$ms Kokkos), almost matching strategy 2 even though we lack the overlapping of the CPU-GPU data transfers.
However, on the AMD node strategy 3 outmatches the other strategies, gaining us a clear speedup over the CPU runs even with $8^3$ sub-grids (which strategy 2 could not provide).
We achieve a runtime of $166$ms with HIP and $189$ms with Kokkos using this strategy on the AMD node.
Overall, strategy 3 provides far better results on the AMD node than strategy 2 and works regardless which kernels (and which GPUs) are used, increasing the portability of our approach of task-based GPU programming.
As noted in the previous section, nothing stops us from using strategy 3 with more underlying GPU executors.
This allows us to combine the advantages of the separate strategies, notably the larger kernels from strategy 3 and the overlapping of data-transfers and kernels from strategy 2.
Indeed, a combination of strategy 2 and 3 usually works best.
In the last two sections of the table, we can see the best parameter combinations we found.

Using $8^3$ sub-grids, we reach runtimes as low as $86.6$ms with CUDA ($93$ms with Kokkos) on the NVIDIA node, or $130$ms with HIP ($148$ms with Kokkos) on the AMD node.
In terms of speedup over the non-aggregated GPU run with this sub-grid size (using $1$ GPU executor and $1$ max aggregated kernel), this results in a speedup of 10.04x for CUDA, 9.35x for Kokkos (on the NVIDIA node), 10.80x for HIP and 11.96x for Kokkos on the AMD node.
Considering that prior to this work, strategy 2 was the state-of-the-art for Octo-Tiger, the speedups of the best combinations over this strategy (with 128 GPU executors) are also worth mentioning: 1.52x for CUDA, 1.57x for Kokkos (on the NVIDIA node), 4.49x for HIP and 5.23x for Kokkos HIP. This highlights that our newly added strategy 3 is already useful on the NVIDIA GPUs, and essential for good performance in Octo-Tiger on the AMD GPUs.
These runtimes per time-step include all the required work, not just the GPU kernel calls and associated data transfers for all sub-grids, but also the ghost cell handling, and other CPU methods such as pre- and post-processing to obtain the final result.
Even for the larger sub-grids, activating an additional strategy seems to work best. On the NVIDIA node, we benefit from using more executors, achieving runtimes as low as $73$ms with CUDA on the NVIDIA node ($81$ms with Kokkos) using $32$ executors. On the AMD node, we still need to combine strategies 2 and 3, getting to runtimes of $101$ms with HIP and $116$ms with Kokkos.

Overall, while all strategies have their own benefits and challenges, there are general trends: \textit{1)} Strategy 3 and 1 work more consistently across AMD and NVIDIA GPUs, \textit{2)} A combination of strategies always works best, and \textit{3)} Kokkos reaches performance close to the native kernels (optimal runs are about $10$\% slower using Kokkos). While this means more optimization is required to close this gap, it is a promising result for the first implementation.


\begin{table*}[p]
  \caption{Setup: Overall number of cells, sub-grids and kernel launches for the utilized (hydro-only) Sedov Blast Wave scenario.}
  \label{results:scenario-size}
  \begin{tabular}{ c  c c c | c c }
      \bottomrule
      \toprule
      \rowcolor{lightgray}
    \multicolumn{6}{c}{\textbf{Blast Wave Scenario with Different Sub-Grid Sizes}} \\
      \rowcolor{lightgray}
    \multicolumn{4}{c}{Grid parameters} & \multicolumn{2}{c}{GPU metrics per time-step} \\
    \midrule
    Sub-grid size & Overall number of cells & Number of (leaf) sub-grids & Ghost cells per sub-grid  & \ \ \ Kernel calls \ \ \ & CPU-GPU data transfers \ \ \ \ \\
    \midrule
    $8^3$ (512) & 262144 & 512 & 2232 & 7680 & 15360 \\
    $16^3$ (4096) & 262144 & 64 & 6552 & 960 & 1920 \\
    \bottomrule
  \end{tabular}
\end{table*}
\begin{table*}[p]
  \caption{Experiment: Runtime per time-step running the Sedov Blast Wave scenario using different work aggregation strategies. The runtime per time-step is the average over 15 time-steps. The scenario is run both on an NVIDIA A100 node and an AMD MI100 node. }
\label{results:times}
  \begin{tabular}{ c  c  c  c | r  r |  r  r }
      \bottomrule
      \toprule
      \rowcolor{lightgray}
      \multicolumn{8}{c}{\textbf{CPU Runs for Comparison [CPU-only]}}            \\
      \rowcolor{lightgray}
      \multicolumn{4}{c}{Performance parameters} & \multicolumn{2}{c}{Runtime per time-step A100 node} & \multicolumn{2}{c}{Runtime per time-step MI100 node} \\
      \midrule
      Cores & Sub-grid size & GPU executors & Max aggregated kernels & Original CPU Impl & Kokkos (CPU) & Original CPU Impl & Kokkos (CPU) \\
      1 & $8^3$ & 0 & 1 & 9068 ms & 10548 ms & 10453 ms & 11140 ms \\
      32 & $8^3$ & 0 & 1 & 521 ms & 462 ms & 518 ms & 530 ms \\
      1 & $16^3$ & 0 & 1 & 6733 ms & 7619 ms & 8380 ms & 8479 ms   \\
      32 & $16^3$ & 0 & 1 & 505 ms &  529 ms & 510 ms & 515 ms  \\
      \bottomrule
      \toprule
      \rowcolor{lightgray}
      \multicolumn{8}{c}{\textbf{Strategy 1: Subdivide Grid into Larger Sub-Grids [GPU-accelerated]}}            \\
      \rowcolor{lightgray}
      \multicolumn{4}{c}{Performance parameters} & \multicolumn{2}{c}{Runtime per time-step A100 node} & \multicolumn{2}{c}{Runtime per time-step MI100 node} \\
      \midrule
      Cores & Sub-grid size & GPU executors & Max aggregated kernels &  CUDA & Kokkos  & HIP & Kokkos  \\
      32 & $8^3$ & 1 & 1 & 869 ms & 870 ms & 1404 ms & 1770 ms \\
      32 & $16^3$ & 1 & 1 & 159 ms & 167 ms & 224 ms & 275 ms \\
      \bottomrule
      \toprule
      \rowcolor{lightgray}
      \multicolumn{8}{c}{\textbf{Strategy 2: Increase the Number of GPU Executors [GPU-accelerated]}}            \\
      \rowcolor{lightgray}
      \multicolumn{4}{c}{Performance parameters} & \multicolumn{2}{c}{Runtime per time-step A100 node} & \multicolumn{2}{c}{Runtime per time-step MI100 node} \\
      \midrule
      Cores & Sub-grid size & GPU executors & Max aggregated kernels &  CUDA & Kokkos  & HIP & Kokkos  \\
      32 & $8^3$ & 1 & 1 & 869 ms & 870 ms & 1404 ms & 1770 ms \\
      32 & $8^3$ & 2 & 1 & 450 ms & 459 ms & 1285 ms & 1526 ms \\
      32 & $8^3$ & 4 & 1 & 245 ms & 261 ms & 849 ms & 1337 ms \\
      32 & $8^3$ & 8 & 1 & 215 ms & 237 ms & 700 ms & 1117 ms \\
      32 & $8^3$ & 16 & 1 & 174 ms & 193 ms & 637 ms & 946 ms \\
      32 & $8^3$ & 32 & 1 & 150 ms & 170 ms & 603 ms & 846 ms \\
      32 & $8^3$ & 64 & 1 & 139 ms & 156 ms & 588 ms & 786 ms \\
      32 & $8^3$ & 128 & 1 & 132 ms & 146 ms & 584 ms & 774 ms \\
      \bottomrule
      \toprule
      \rowcolor{lightgray}
      \multicolumn{8}{c}{\textbf{Strategy 3: Increase the Maximum Number of On-The-Fly Aggregated Kernels [GPU-accelerated]}}            \\
      \rowcolor{lightgray}
      \multicolumn{4}{c}{Performance parameters} & \multicolumn{2}{c}{Runtime per time-step A100 node} & \multicolumn{2}{c}{Runtime per time-step MI100 node} \\
      \midrule
      Cores & Sub-grid size & GPU executors & Max aggregated kernels &  CUDA & Kokkos  & HIP & Kokkos  \\
      32 & $8^3$ & 1 & 1 & 869 ms & 870 ms & 1404 ms & 1770 ms \\
      32 & $8^3$ & 1 & 2 & 485 ms & 492 ms & 735 ms & 934 ms \\
      32 & $8^3$ & 1 & 4 & 278 ms & 294 ms & 414 ms & 512 ms \\
      32 & $8^3$ & 1 & 8 & 207 ms & 203 ms & 253 ms & 311 ms \\
      32 & $8^3$ & 1 & 16 & 168 ms & 181 ms & 188 ms & 224 ms \\
      32 & $8^3$ & 1 & 32 & 153 ms & 154 ms & 164 ms & 191 ms \\
      32 & $8^3$ & 1 & 64 & 158 ms & 152 ms & 160 ms & 175 ms \\
      32 & $8^3$ & 1 & 128 & 156 ms & 165 ms & 166 ms & 189 ms \\
      \bottomrule
      \toprule
      \rowcolor{lightgray}
      \multicolumn{8}{c}{\textbf{Best combinations of Strategy 2 and 3 for the sub-grid size $8^3$ [GPU-accelerated]}}            \\
      \rowcolor{lightgray}
      \multicolumn{4}{c}{Performance parameters} & \multicolumn{2}{c}{Runtime per time-step A100 node} & \multicolumn{2}{c}{Runtime per time-step MI100 node} \\
      \midrule
      Cores & Sub-grid size & GPU executors & Max aggregated kernels &  CUDA & Kokkos  & HIP & Kokkos  \\
      32 & $8^3$ & 64 & 8 & 87.4 ms & \textbf{93 ms} & 142 ms & 176 ms \\
      32 & $8^3$ & 128 & 8 & \textbf{86.6 ms} & 94 ms & 140 ms & 175 ms \\
      32 & $8^3$ & 128 & 16 & 92 ms & 96 ms & \textbf{130 ms} & 149 ms \\
      32 & $8^3$ & 128 & 32 & 99 ms & 102 ms & 131 ms & \textbf{148 ms} \\
      \bottomrule
      \toprule
      \rowcolor{lightgray}
      \multicolumn{8}{c}{\textbf{Best combinations of Strategy 2 and 3 for the sub-grid size $16^3$ [GPU-accelerated]}}            \\
      \rowcolor{lightgray}
      \multicolumn{4}{c}{Performance parameters} & \multicolumn{2}{c}{Runtime per time-step A100 node} & \multicolumn{2}{c}{Runtime per time-step MI100 node} \\
      \midrule
      Cores & Sub-grid size & GPU executors & Max aggregated kernels &  CUDA & Kokkos  & HIP & Kokkos  \\
      32 & $16^3$ & 32 & 1 & \textbf{73 ms} & 85ms & 123 ms & 146 ms \\
      32 & $16^3$ & 32 & 2 & 75 ms & \textbf{81 ms} & 103 ms & 124 ms \\
      32 & $16^3$ & 32 & 4 & 78 ms & 83ms & \textbf{101 ms} & \textbf{116} ms \\
      \bottomrule
  \end{tabular}
\end{table*}

\section{Conclusion and Future work}
\label{sec:conclusion}
All three work aggregation strategies discussed can successfully increase the workload, and thus the utilization, of the GPU. However, the strategies come with their own respective benefits and challenges: 

\textit{1)} Using larger sub-grids in Octo-Tiger works well to improve the GPU performance and to reduce the number of ghost cells. But this is an application-specific way of increasing the workload per GPU kernel. Furthermore, it has implications on the mesh refinement and the gravity solver.

\textit{2)} Using multiple GPU executors to run the GPU kernels concurrently is an application-independent strategy that can work with any HPX application. It provides additional benefits as we can overlap CPU-GPU data transfers with GPU kernel executions. Our results show that it is, however, vendor-specific. While the strategy works well on the NVIDIA GPU, we struggle on the AMD GPU to even get close to the CPU performance.

\textit{3)} Our newly introduced work aggregation executors perform well across the tested platforms. This strategy is even application-independent. However, it requires the programmer to mark kernels that are compatible to be aggregated.

Overall, our results show that a mixture of all strategies works best.
Exactly what mixture to use depends on the hardware and the scenario we use:
\textit{1) Hardware}: How well does the GPU/runtime API handle the respective strategy? We have seen for strategy 2 that there can be vast differences depending on the GPU, even though they should have similar performance.
\textit{2) Scenario}: How feasible is it to actually move to larger sub-problems? If we ever encounter a scenario with Octo-Tiger where larger sub-grids provide a better trade-off, we would want to add this to our mix of aggregation strategies.

Overall, this work provides clearly improves stellar simulations with Octo-Tiger. The new work aggregation strategy yields significant speedups on all devices for the hydro solver using the current state-of-the-art sub-grid size of 8x8x8. 


For future work, we are currently preparing Octo-Tiger scalability runs on both Perlmutter and Fugaku. We plan to use the new Kokkos kernels on both machines and, given the node-level performance impact shown in this work, the new explicit work aggregation strategy to steer the workload per kernel depending on whether it runs on a A100 GPU on Perlmutter or on one A64FX CPU core during a run on Fugaku.

\section*{Copyright notice}
\textcopyright 2022 IEEE. Personal use of this material is permitted.
  Permission from IEEE must be obtained for all other uses, in any current or future 
  media, including reprinting/republishing this material for advertising or promotional 
  purposes, creating new collective works, for resale or redistribution to servers or 
  lists, or reuse of any copyrighted component of this work in other works. 

\bibliographystyle{IEEEtran}
\bibliography{ref.bib}

\end{document}